\documentclass[twocolumn,english,aps,prl,reprint,superscriptaddress]{revtex4-2}
\usepackage{amsmath}
\usepackage[utf8]{inputenc}
\usepackage{txfonts}
\usepackage{amssymb}
\usepackage{graphicx}
\usepackage{placeins}
\usepackage{booktabs}
\usepackage{xcolor}
\usepackage{mathrsfs}
\usepackage{babel}
\usepackage{microtype}
\usepackage[colorlinks=true, urlcolor=blue, linkcolor=blue, citecolor=blue]{hyperref}

\renewcommand{\selectlanguage}[1]{}
\def\eV{{\,\mathrm{eV}}}
\def\meV{{\,\mathrm{meV}}}

\begin{document}
\title{Charge disproportionation driven polar magnetic metallic double-layered perovskite $\mathrm{Sr}_3\mathrm{Co}_2\mathrm{O}_7$ }

\author{Hong-Fei Huang}
\affiliation{School of Physical Science and Technology, 
Soochow University, Suzhou 215006, China}
\author{Houssam Sabri}
\affiliation{Department of Physics and Astronomy, University of New Hampshire,
Durham, New Hampshire 03824, USA}
\author{Jiadong Zang}
\email{jiadong.zang@unh.edu}
\affiliation{Department of Physics and Astronomy, University of New Hampshire,
Durham, New Hampshire 03824, USA}
\author{Jie-Xiang Yu}
\email{jxyu@suda.edu.cn}
\affiliation{School of Physical Science and Technology, 
Soochow University, Suzhou 215006, China}
\begin{abstract}
Strong coupling among spontaneous structural symmetric breaking, magnetism and metallicity in an intrinsic polar magnetic metal 
can give rise to novel physical phenomena and holds great promise for applications in spintronics. 
Here, we elucidate the mechanism of metallic ferroelectricity in the recently discovered polar metal $\mathrm{Sr}_3\mathrm{Co}_2\mathrm{O}_7$.
Our first-principles calculations reveal that both the spontaneous ferroelectric displacements and the metallicity originate from charge disproportionation of Co ions. 
This is characterized by an inverted ligand-field splitting of the Co $t_{2g}$ orbitals at one site, while the metallic behavior is preserved by the $t_{2g}$ orbitals at both sites.
The charge disproportionation stabilizes the asymmetric phase Within the framework of the on-site Hubbard $U$ interaction. 
We thus propose that in related transition metal oxides, charge disproportionation within specific orbitals can concurrently drive metallicity and ferroelectricity, enabling strong coupling between these properties.
More remarkably, this mechanism allows for the coexistence of magnetism, as evidenced in $\mathrm{Sr}_3\mathrm{Co}_2\mathrm{O}_7$. 
Our findings highlight a promising avenue for realizing polar magnetic metals and provide a new design principle for exploring multifunctional materials.
\end{abstract}

\maketitle

Ferroelectricity arisen from spontaneous structural symmetry breaking has received broad interest from both fundamental physics and functional application perspectives. 
While spontaneous polarization naturally emerges in insulators \cite{devonshire_theory_1954}, its existence in metals is elusive due to screening by conduction electrons, 
which typically quenches any net polarization \cite{anderson_symmetry_1965}.
Nevertheless, noncentrosymmetric magnetic materials provide a fertile ground for novel spin textures and emergent transport phenomena\cite{muhlbauer_skyrmion_2009,yu_real-space_2010, neubauer_topological_2009, yin_topological_2015, hou_thermally_2017, vistoli_giant_2018, shao_topological_2019, yu_thermally_2019}. 
Therefore, magnetic polar metals-- a new state of matter integrating ferroelectricity, magnetism, and metallicity--exhibit distinct magnetoelectric characteristics. 
Unlike insulating multiferroics, they enable the exploration of exotic electronic states derived from the metallic magnetoelectric coupling within a new paradigm.
Consequently, to avoid the screening effect, people employ the weak coupling principle emerges as a key design strategy, i.e., 
to isolate ferroelectric displacement and metallic behavior in distinct layers or sublattices.
It has been demonstrated in heterostructure systems\cite{kim_polar_2016,wetherington_2-dimensional_2021, deng_polarization-dependent_2024}, 
metallic doping in ferroelectric semiconductors\cite{kolodiazhnyi_persistence_2010}, 
and some intrinsic polar magnetic metals\cite{zabalo_switching_2021,zhang_correlated_2024}.
However, this strategy limits the coupling between lattice, orbital and spin, 
so that breaking through the constraint of the weak coupling principle is essential to design a strong-coupled polar magnetic metal material.
 
Charge disproportionation, often occurring in transition metal oxides such as 
ferrites\cite{takano_charge_1977, woodward_structural_2000}, 
cobaltites\cite{lee_disproportionation_2005} 
and nickelates\cite{torrance_systematic_1992, alonso_charge_1999, mazin_charge_2007}, 
is an ubiquitous mechanism for generating spontaneous structural inhomogeneity. 
Whereas the electric-driven Jahn-Teller distortion only entails on-site rearrangement of electron orbitals, 
charge disproportionation involves nominal charge transfer between neighboring lattice sites
induced by the fundamental reason of the instability of the average valence state in Fe$^{4+}$, Co$^{4+}$ and Ni$^{3+}$.
On the common ground of broken inversion symmetry, it is expected that charge disproportionation is able to drive ferroelectricity in corresponding transition metal oxides.
In this letter, we discovered a realization of this mechanism in  $\mathrm{Sr}_3\mathrm{Co}_2\mathrm{O}_7$, 
a newly found double-layered perovskite-based magnetic polar metal. \cite{zhou_geometry-driven_2025}
Based on first-principles calculations, the asymmetric phase with ferroelectric displacement is energetically favorable over the symmetric phase.  
Distinct electron occupancies and ligand field of the localized $t_{2g}$ orbitals are identified in two Co sites. 
Such charge disproportionation preserves the metallic property and reduces the total energy with the influence of on-site electron-electron Coulomb repulsion. 
The magnetic properties are also discussed, 
and the calculated results are in good agreement with the experimental findings.

The calculations were performed by using the Vienna ab initio simulation (VASP) package\cite{kresse_efficiency_1996,kresse_efficient_1996} based on the density-functional theory (DFT) with the projector augmented-wave (PAW) pseudo-potentials~\cite{blochl_projector_1994,kresse_ultrasoft_1999}.
The generalized gradient approximation in Perdew–Burke–Ernzerhof (PBE) \cite{perdew_generalized_1996} formation was used as the exchange-correlation energy.
To include the strong-correlation effects of the localized $3d$ electrons, 
we employed the Hubbard $U$ method\cite{anisimov_band_1991} of $U=4.0\eV$ and $J=0.9\eV$ on Co orbitals.
The plane-wave energy cutoff was set to be 550 eV, and the Monkhorst-Pack $20\times20\times4$  k-mesh was employed.
The experimental lattice constants as $a=b=3.868$~\AA~and $c=19.80$~\AA~was used throughout the calculations.
The structures for each structural and magnetic configuration were optimized 
until the Hellmann-Feynman forces on all atoms were less than
$1\meV/$\AA. 
Subsequently, we constructed a tight-binding Hamiltonian by performing a unitary transformation from the plane-wave basis to Wannier functions (WFs) using the band disentanglement method\cite{souza_maximally_2001} implemented in the Wannier90 package\cite{mostofi_updated_2014} to construct the tight-binding Hamiltonian. 
Co($3d$) and O($2p$) orbitals were chosen for the projection. 
The on-site energies and electron occupancies for these atomic orbitals can be obtained.


\begin{figure}
\includegraphics[width=\columnwidth]{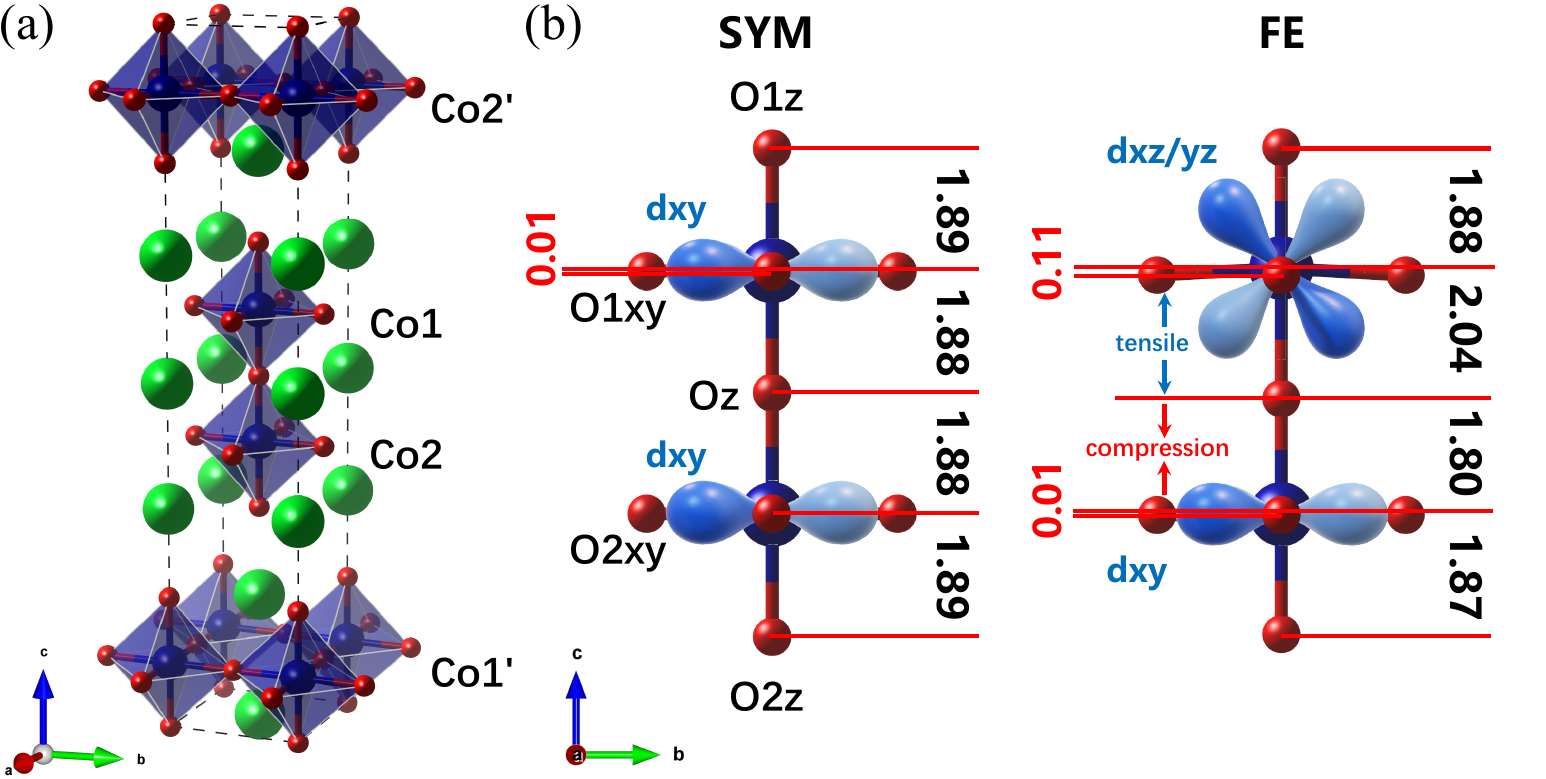}
\caption{(a) a $1\times1\times1$ tetragonal unit cell of Sr$_3$Co$_2$O$_7$. 
Co1/Co2 correspond to the two Co sites in one double-layered structure respectively,
and Co1' and Co2' are the corresponding Co sites in the other double-layered structure.
In (a), Co1/Co2, 
(b) The inter-atomic distances along the $c$-axis for the symmetric (SYM) and asymmetric ferroelectric (FE) phases respectively in units of \AA.
Oz, O1z/O2z, and O1xy/O2xy correspond to  
the apical oxygen connecting Co1 and Co2,
the apical oxygen atoms connecting Co1/Co2 respectively,
and equatorial oxygen atoms in the CoO$_2$ layers centered by Co1/Co2, respectively. 
Red numbers indicate the displacements between Co1/Co2 and O1xy/O2xy within the same CoO$_2$ layer.
The occupied $d_{xy}$ and $d_{xz/yz}$ orbitals, showing the charge disproportionation in FE phase are indicated.
}

\label{fig:structure} 
\end{figure}

Fig.~\ref{fig:structure}(a) shows the $1\times1\times1$ tetragonal unit cell of double-layered perovskite Sr$_3$Co$_2$O$_7$.
Each pair of corner-shared CoO$_6$ octahedron along $c$-axis forms a primitive unit, 
in which the central magnetic cobalt atoms are labeled as Co1 and Co2 respectively.
Our structural optimization yields centrosymmetric (SYM) and asymmetric (FE) phases shown in Fig~\ref{fig:structure}(b).
In the SYM phase, Co1 and Co2 are equivalent and the apical oxygen, labeled as Oz, sitting in between Co1 and Co2 is the inversion center.
The FE phase exhibits a significant displacement of Co1 atoms along the $c$-axis, causing the ferroelectricity. Co2 atoms have negligible displacements, thus do not contribute to the polarization. 
The supporting calculations in a expanded $2\times2\times1$ tetragonal supercell were also performed and show almost no tilting and rotation in the CoO$_6$ octahedra. 
That is consistent with the x-ray diffraction data of the experiment.
Two magnetic configurations were considered.
One is the ferromagnetic (FM) state and the other is the A-type antiferromagnetic (A-AF) state where the interlayer Co1-Co2 coupling is antiferromagnetic and intralayer Co1-Co1 or Co2-Co2 coupling is ferromagnetic.
The relative total energies for the two structural phases and two magnetic orderings are listed in Table.\ref{Table:energy}.
With either FM or A-AF magnetic ordering, FE phase has lower energy than SYM phase.
The energy difference of about $35\meV$ per Co indicates the transition temperature of the FE phase is higher than the room temperature. Structural transition is indeed not observed below the room temperature experimentally.
Furthermore, the FM ordering in the FE phase is about $10\meV$ per Co lower in energy than the A-AF ordering, 
indicating a weak ferromagnetic ground state with a Curie temperature below room temperature.
These conclusions remain unchanged when varying the Hubbard $U$ parameter from $4.0$ to $6.0\eV$.

\begin{table}
\caption{Relative total energies, total magnetization per Co, and local magnetic moments for the Co1 and Co2 atoms for states with centrosymmetric/asymmetric (SYM/FE) and ferromagnetic/A-type antiferromagnetic (FM/A-AF) phases. The total energies per Co are in relative to the non-polar centrosymmetric ferromagnetic (SYM+FE) phase.}
\begin{ruledtabular}
\begin{tabular}{lrrrr}
 & $E$(meV) & $m_\mathbf{tot}$($\mu_{B}$) & $m_\mathbf{Co1}$($\mu_{B}$) & $m_\mathbf{Co2}$($\mu_{B}$) \\
\hline 
SYM+FM   &   0.0 & 1.97 &  1.71 & 1.71 \\
SYM+A-AF &  80.6 & 0.00 & -1.64 & 1.64 \\ 
FE+FM    & -35.0 & 1.92 &  1.32 & 2.00 \\
FE+A-AF  & -21.5 & 0.62 & -1.23 & 2.02 \\ 
\end{tabular}
\end{ruledtabular}
\label{Table:energy}
\end{table}

The structural asymmetry is further reflected in the local magnetic moments of Co1 and Co2.
While the total magnetization is nearly identical between the SYM+FM and FE+FM phases, 
the local moments differ significantly: in the FE+FM phase, Co1 and Co2 have moments of $1.32\mu_B$ and $2.00\mu_B$, respectively, 
compared to identical moments of $1.71\mu_B$ in the SYM+FM phase.
This indicates distinct spin and valence states for Co1 and Co2 in the polar phase.

\begin{figure}[t]
\includegraphics[width=\columnwidth]{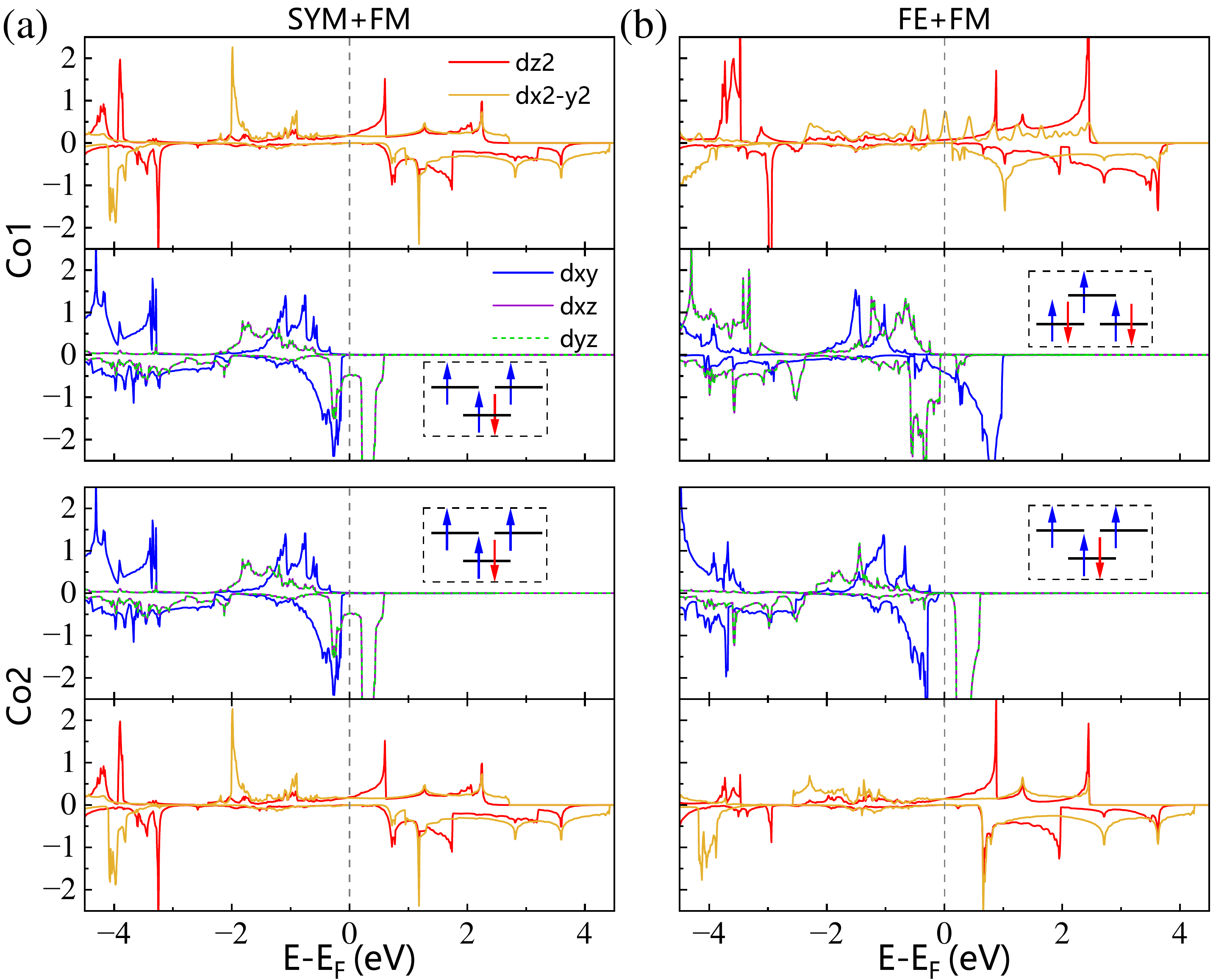}
\caption{The projected density of states (PDOS) of Co($3d$) orbitals, including $t_{2g}$ ($d_{xz}, d_{yz}$ and $d_{xy}$) and $e_g$ ($d_{z^2}$ and $d_{x^2-y^2}$) orbitals for the (a) SYM+FM phase (b) FE+FM phase. 
The Fermi energy is set to zero.
The energy levels and electron filling of $t_{2g}$ orbitals for Co1 and Co2 in both cases is shown schematically.}
\label{fig:pdos_z} 
\end{figure}

To provides valuable insights into the stable FE phase, we investigate 
the electronic structure of Co ions.
Fig.~\ref{fig:pdos_z} gives the projected density of states (PDOS) of Co atoms in SYM and FE phases in FM ordering.
No significant difference is observed for $e_g$ orbitals between SYM and FE phases. Even in FE phase, $e_g$ occupations in Co1 and Co2 are not visibly different.
For both phases, the $e_g$ orbitals, composed of $d_{3z^2-r^2}$ and $d_{x^2-y^2}$, are partially occupied and have non-zero PDOS at the Fermi level in both spin channels.
These partially occupied orbitals, especially the strongly bonded $d_{x^2-y^2}$, are responsible for metallicity in Sr$_3$Co$_2$O$_7$.

In contrast, the occupancies of the $t_{2g}$ orbitals are significantly different for Co1 and Co2 in the FE phase.
With the tetragonal ligand field, the degenerate $t_{2g}$ orbitals are further lifted into singlet $d_{xy}$ and doublet $d_{xz}$ and $d_{yz}$.
In the spin majority channel, all three $t_{2g}$ orbitals are occupied for all Co's in both SYM and FE phases.
In the spin minority channel, both phases behave differently.
In the SYM phase, major peaks of the degenerate $d_{xz}/d_{yz}$ orbitals are located just above the Fermi level,
while that of the $d_{xy}$ orbital has a lower energy and is just below the Fermi level.
However, in the FE phase, the band ordering is inverted in Co1, 
where the degenerate $d_{xz}/d_{yz}$ orbitals is mainly occupied and the $d_{xy}$ orbital has a higher energy and is located above the Fermi level. 
Ordering of Co2 is still the same as the SYM phase.

Considering the nominal +2 valence of Sr and -2 for O, each Co atom in the SYM phase has an average valence of +4, corresponding to a $d^5$ configuration.
Its intermediate spin state has the configuration $t_{2g}^4e_{g}^{2}\underline{L}$, 
where \underline{L} denotes a ligand hole is suggested in perovskite ferromagnet SrCoO$_3$ with the same valence state\cite{potze_possibility_1995}. 
Taking spin configuration into account, Co in SYM phase thus has $t_{2g}(\uparrow^3\downarrow^1)e_{g}^2\underline{L}$.
In the FE phase, the unpolarized Co2 remains the same as that in the SYM phase,
while the polarized Co1 has the spin state $t_{2g}(\uparrow^3\downarrow^2)e_{g}^2\underline{L}^2$
where $t_{2g}(\downarrow^2)$ denotes the occupied $d_{xz\downarrow}/d_{yz\downarrow}$ orbitals.
Such intermediate spin state of Co is similar to that in the perovskite LaCoO$_3$\cite{korotin_intermediate-spin_1996}.
Therefore, Co1 in the FE phase exhibits one more formal electron than Co2.

The charge disproportionation of Co in the FE phase is coupled with the ferroelectric displacement.
Fig.\ref{fig:structure}(b) shows the interatomic distances along $c$-axis.
In the FE phase, the Co1–Oz distance elongates while the Co2–Oz distance shortens. 
This distortion inverts the ligand-field splitting for Co1, 
lowering the energy of the $d_{xz}/d_{yz}$ orbitals relative to $d_{xy}$, 
which is opposite to the splitting in Co2. 
The rearrangement of the $t_{2g}$ orbital energies in Co1 thus underpins the simultaneous electronic and structural asymmetry.

\begin{figure}[t]
\includegraphics[width=\columnwidth]{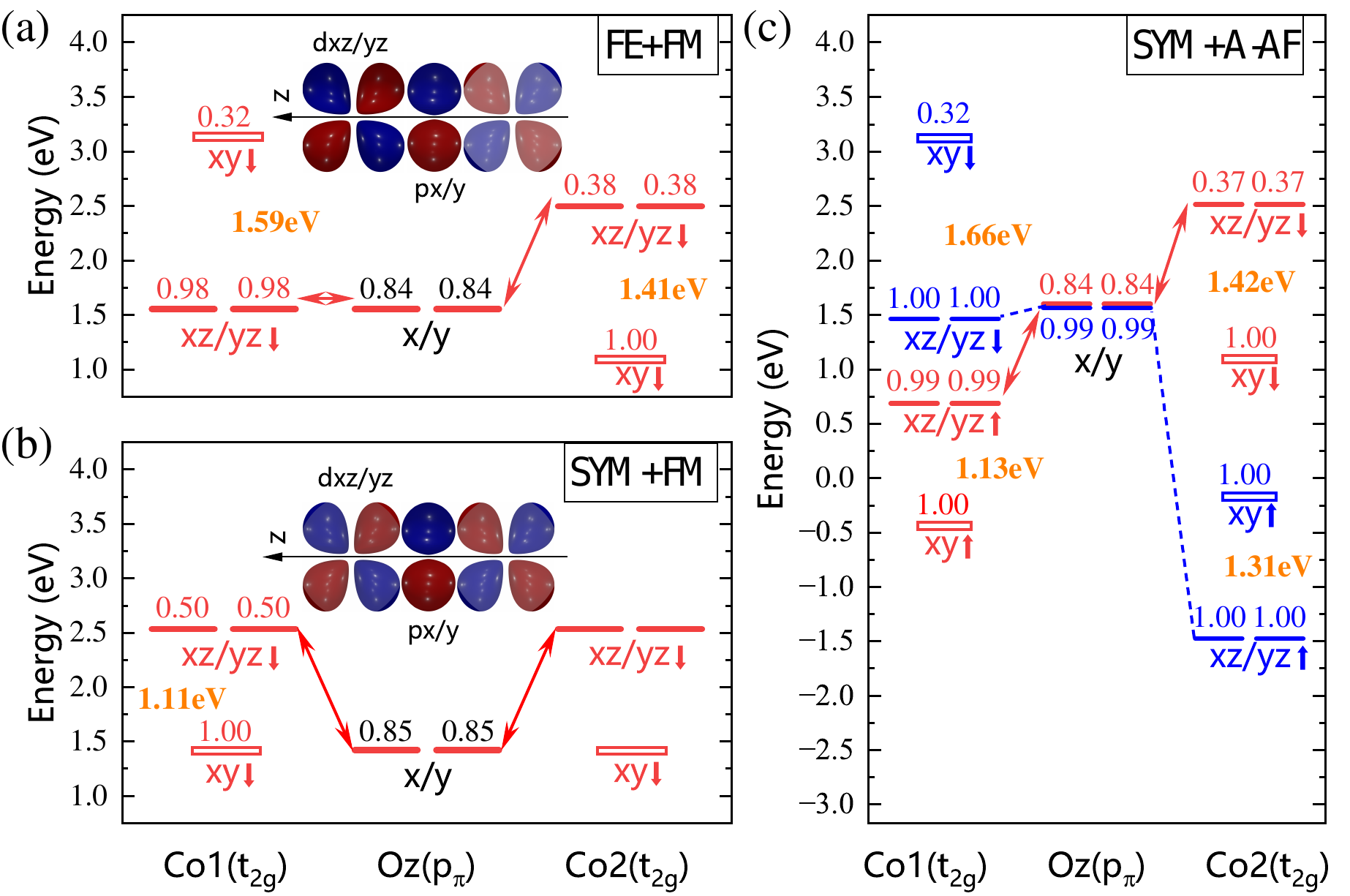}
\caption{The on-site energy levels (lines) and electron occupancies (numbers) of the atomic orbitals associated with polarization [Co1($t_{2g}$), Oz($p_{\pi}$), Co2($t_{2g}$)] for the (a) FE+FM phase (b) SYM+FM phase (c) FE+A-AF phase. 
Red and blue signifies the spin down and spin up channel, respectively. 
$\uparrow$ and $\downarrow$ refers to spin majority and minority relative to Co's local moment.
Double-arrows indicate the virtual exchange pathways via $pd\pi$ bonding.
Dashed lines in (c) indicate the exchange pathways with all orbitals fully occupied.}

\label{fig:on-site} 
\end{figure}

Both the SYM and FE phases are dynamically stable without soft modes,
so that the displacement of the apical oxygen atom Oz between Co1 and Co2 away from the centrosymmetric position has a cost of elastic energy.
This elastic energy is estimated as $15\meV$ per Co based on the finite displacement approximation (See Supplementary Materials for details).
In order to stabilize the FE phase, the energy gain from charge disproportionation  must be significant.
Using the WF-based tight-binding Hamiltonian, 
we have obtained the on-site energy levels of the Co($t_{2g}$) orbitals and the $p_{x}/p_{y}$ orbitals of Oz. 
Meanwhile, by projecting and integrating the eigenstates after diagonalizing the Hamiltonian, 
we have obtained the electron occupancies of each orbital.
Fig.\ref{fig:on-site}(a)(b) give the results for both the SYM+FM and FE+FM phases in spin down channels because the $t_{2g}$ orbitals in the spin up channel are fully occupied.
In the SYM phase, the degenerate $d_{xz\downarrow}/d_{yz\downarrow}$ orbitals are about $1.1\eV$ higher in energy than the fully occupied $d_{xy\downarrow}$ orbital. 
Their occupancies about 0.5 is the consequence of the hybridization with $2p$ orbitals of oxygen ligands, 
so that nominally we still regard the molecular-like $d_{xz\downarrow}/d_{yz\downarrow}$ orbitals as the unoccupied state.
In the FE phase, it is clear that for Co1, the fully occupied $d_{xz\downarrow}/d_{yz\downarrow}$ orbitals are about $1.6\eV$ lower than the $d_{xy\downarrow}$ orbital,
whose occupancy of 0.32 indicates unoccupied state.
Meanwhile, the energy splitting between $d_{xz\downarrow}/d_{yz\downarrow}$ and $d_{xy\downarrow}$ for Co2 is about $1.4\eV$, which is larger than that in the SYM phase.
Both the inverse ligand field splitting in the Co1($t_{2g}\downarrow$) and the enlarged splitting in the Co2($t_{2g}\downarrow$) are consist with the PDOS and the structural results.

\begin{table}
\caption{The occupancies of the relevant Co($3d$) orbitals, 
	the potential difference $\Delta V = V_{xz(yz)\downarrow} - V_{xy\downarrow}$ 
	and the single particle Hubbard $U$ energies $E^{(U)}$ of three $t_{2g}\downarrow$ in SYM+FM and FE+FM phase, respectively.}
\begin{ruledtabular}
\begin{tabular}{lrrrrrr}
 & $n_{\uparrow}$ & $n_{\downarrow}$ & $n_{xz(yz)\downarrow}$ & $n_{xy\downarrow}$ & $\Delta V$ (eV) & $E^{(U)}$ (eV) \\
\hline 
Co in SYM  &  4.33 & 2.62 &  0.50 &  1.00 &  1.55 & 0.511 \\
Co1 in FE  &  4.16 & 2.85 &  0.98 &  0.32 & -2.05 & 0.297 \\ 
Co2 in FE  &  4.41 & 2.46 &  0.38 &  1.00 &  1.95 & 0.602
\end{tabular}
\end{ruledtabular}
\label{Table:n}
\end{table}

Based on these results, we estimate the electronic energy gain from charge disproportionation within the DFT+$U$ framework.
The single particle potential as a function of the electron occupancy $n_{m\sigma}$ for each orbital with orbit index $m$ and spin index $\sigma$ is given by\cite{anisimov_band_1991}
\begin{eqnarray}
V_{m\sigma} & = & U\sum_{m'}\left(n_{m'-\sigma}-n_0\right) + (U-J)\sum_{m'\ne m}\left(n_{m'\sigma}-n_0\right) \\
            & = & -J n_{\sigma} - (U-J) n_{m\sigma} + \frac{1}{10} (U+4J) n_{d}
\label{eq:u}
\end{eqnarray}
where $n_{\sigma}=\sum_{m}n_{m\sigma}$ is the total occupancy number in the spin channel $\sigma$, 
$n_{d}=n_{\sigma}+n_{-\sigma}=\sum_{m,\sigma}n_{m\sigma}$ is the total occupancy for all $d$ orbitals
and $n_{0}=n_{d}/10$ is the average occupancy for each $d$ orbital.
For orbitals in the same spin channel, the larger of $n_{m\sigma}$, the lower the potential of this orbital.
Using the occupancies listed in Table \ref{Table:n}, we computed $V_{m\downarrow}$ for the three $t_{2g}$ orbitals on each Co site.
Then the potential difference $\Delta V$ between $d_{xz\downarrow}/d_{yz\downarrow}$ and $d_{xy\downarrow}$ 
and the single particle Hubbard $U$ energies $E^{(U)}=\sum_{m} V_{m\downarrow} n_{m\downarrow}$ were also calculated and listed in  Table \ref{Table:n}.
As a result, the magnitude of $\Delta V$ for Co2 in the FE phase is about 0.4$\eV$ larger than in the SYM phase, 
consistent with the on-site energy results in Fig.\ref{fig:on-site}.
Moreover, $E^{(U)}(\textrm{Co}_\textrm{SYM}) - 1/2 [ E^{(U)}(\textrm{Co1}_\textrm{FE}) + E^{(U)}(\textrm{Co2}_\textrm{FE})] \approx 60\meV$, which means the FE phase has about $60\meV$ per Co lower in electronic Hubbard $U$ energy than the SYM phase.
After subtracting the elastic energy cost of $\sim 15\meV$ per Co, 
we obtain a net energy gain of $\sim 45\meV$ per Co for the FE phase, 
in excellent agreement with the total energy difference in Table \ref{Table:energy}.

\begin{table}[b]
\caption{Magnetic exchange parameters and the corresponding distances between Co atoms labeled in Fig.\ref{fig:structure}.}
\begin{ruledtabular}
\begin{tabular}{crrrr}
 &Co1-Co1&Co2-Co2& Co1-Co2&Co1-Co2'\\
 \hline 
J (meV)&12.6&22.8& 4.9&0.0\\ 
 d (\AA) &3.868 &3.868 & 3.842 &5.452 \\\end{tabular}
\end{ruledtabular}
\label{Table:exchange}
\end{table}

It is obvious that the charge disproportionation works for both FM and A-AF spin ordering 
as it primarily involves on-site interactions.
The partially occupied itinerant $e_g$ electrons typically favor ferromagnetism via $pd\sigma$ exchange.
Here we will show that the localized $t_{2g}$ electrons also promote FM ordering.
In Fig.\ref{fig:on-site}(a) and (c), the on-site energies and the corresponding occupancies of the apical Oz($px/py$) orbitals are displayed for FM and A-AF ordering respectively.
In FM phase and A-AF phase, the only exchange pathway that affects the total energy via $pd\pi$ bonding between Co1 and Co2 is Co1($d_{xz}/d_{yz}\downarrow$)-Oz($p_{x}/p_{y}$)-Co2($d_{xz}/d_{yz}\downarrow$) 
and Co1($d_{xz}/d_{yz}\uparrow$)-Oz($p_{x}/p_{y}$)-Co2($d_{xz}/d_{yz}\downarrow$) respectively,
where $\uparrow$ and $\downarrow$ correspond to the spin majority and minority relative to the local magnetic moments on Co.
Under the exchange models involving the corresponding orbitals, the energy reduction for both spin ordering phases is obtained by the perturbation theory (See Supplementary Materials for the details.). 
As a result, the energy difference between FM and A-AF phase is on the order of $(t_1^2 t_2^2/U^{\prime4})\Delta_s1$,
where $t_1$ and $t_2$ refer to the hopping of Co1($d_{xz}/d_{yz}$)-Oz($p_{x}/p_{y}$) and Oz($p_{x}/p_{y}$)-Co2($d_{xz}/d_{yz}$) respectively, 
$U^{\prime}$ is the effective Hubbard $U$ for Co(3d) electrons,
and the $\Delta_s$ is the spin splitting of Co1($d_{xz}/d_{yz}$) orbitals.
$\Delta_s$ is positive because $d_{xz}/d_{yz}$ in the spin majority has lower energy than in the spin minority. 
In other words, within the localized limit $U\gg t_1, t_2$, the FM phase generate more energy than the A-AF phase as long as the spin splitting is positive.
This analysis is supported by our calculated exchange parameters (Table \ref{Table:exchange}), obtained using the TB2J package, 
which show ferromagnetic coupling for all significant interactions (in-plane Co1–Co1, Co2–Co2, and out-of-plane Co1–Co2).

\begin{figure}[t]
\includegraphics[width=\columnwidth]{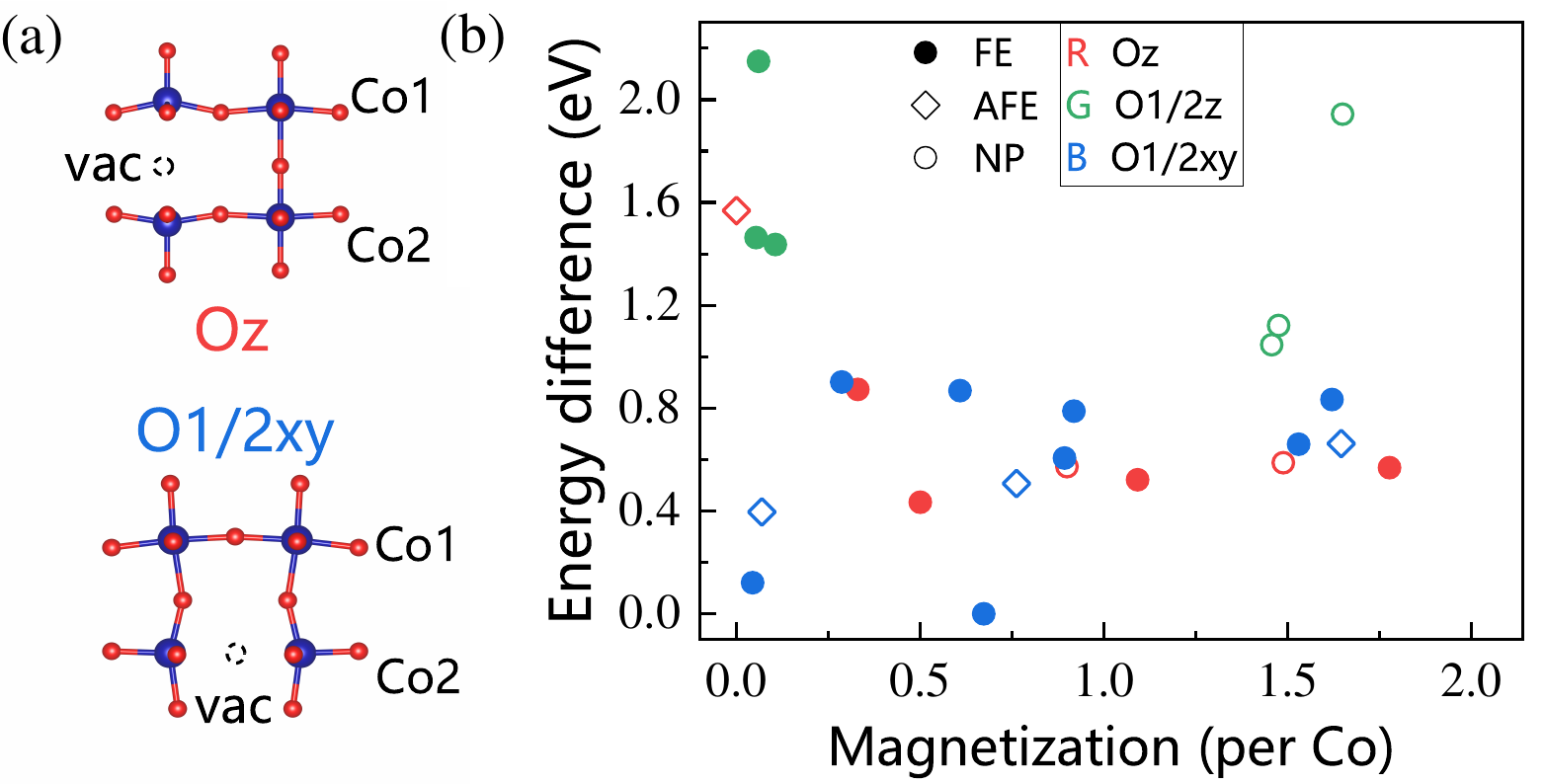}
\caption{The relative total energies (per Co) and the corresponding net magnetization of structures under different locations of the oxygen vacancies labeled in Fig.\ref{fig:structure}. 
Different colors represent vacancies at different locations. 
Hollow circles, solid circles and hollow diamonds denote non-polar (NP), FE and anti-ferroelectric (AFE) phase respectively.}
\label{fig:vac} 
\end{figure}

These ab-initio results are consistent with the experimental observation. 
\cite{zhou_geometry-driven_2025}
At temperatures below 100K, the rise of remanence indicates the ferromagnetic state. 
Although an antiferromagnetic order resembling altermagnetic behavior is stabilized at intermediate temperature between 100K and 200K, the polar state is truely robust up to the room temperature. 
However, the low temperature saturation magnetization is about $1.2\mu_B$ per Co, 
which is significantly lower than the calculated value of $1.92\mu_B$ per Co.
This is likely attributed to the effects of impurities. 
To explain this discrepancy, we performed total energy calculations in a $2\times2\times1$ supercell (32 Co atoms) with a single oxygen vacancy at various locations, 
corresponding to the chemical formula Sr$_3$Co$_2$O$_{6.875}$. 
Fig.\ref{fig:vac} shows the energy results and net magnetization for configurations with various locations of the vacancy in non-polar (NP), ferroelectric and anti-ferroelectric (AFE) phases respectively.
The non-centrosymmetric FE phase still has the lowest total energy while the net magnetization for that state is only about $0.7\mu_B$.
The reason is that with the formation of an oxygen vacancy, 
the nearby Co atoms gains electrons and adopts a 3+ valence state.
Whether through direct exchange (without O mediation) or superexchange (via O atoms), the exchange interaction between two Co$^{3+}$ cations always results in local antiferromagnetic coupling.
Thus, a tiny number of oxygen vacancy defects can greatly suppress the overall magnetization, while they also allow the FE phase to continue to exist stably.

Unlike the charge disproportionation in nickelates with metal-to-insulator transitions\cite{torrance_systematic_1992,alonso_charge_1999, mazin_charge_2007}, 
that in Sr$_3$Co$_2$O$_{7}$ is confined to the $t_{2g}$ electrons and does not involve the $e_g$ electrons,
so the metallicity is robust.
More importantly, the PDOS results shows both $e_g$ and $t_{2g}$ orbitals crossing the Fermi level so that the $t_{2g}$ orbitals hold both the asymmetric and metallic features.
The strong coupling between the ferroelectric displacement and metallic behavior is expected in Sr$_3$Co$_2$O$_{7}$.

H.-F. H. and J.-X. Y. was supported by the National Natural Science Foundation of China (12274309).
\bibliography{main}

\end{document}